# Transferable Adversarial Attacks on Audio Deepfake Detection

Muhammad Umar Farooq, Awais Khan, Kutub Uddin and Khalid Mahmood Malik
College of Innovation & Technology, University of Michigan-Flint, MI, 48502, USA
mufarooq@umich.edu, mawais@umich.edu, kutub@umich.edu, drmalik@umich.edu

## Abstract

*Audio deepfakes pose significant threats, including impersonation, fraud, and reputation damage. To address these risks, audio deepfake detection (ADD) techniques have been developed, demonstrating success on benchmarks like ASVspoof2019. However, their resilience against transferable adversarial attacks remains largely unexplored. In this paper, we introduce a transferable GAN-based adversarial attack framework to evaluate the effectiveness of state-of-the-art (SOTA) ADD systems. By leveraging an ensemble of surrogate ADD models and a discriminator, the proposed approach generates transferable adversarial attacks that better reflect real-world scenarios. Unlike previous methods, the proposed framework incorporates a self-supervised audio model to ensure transcription and perceptual integrity, resulting in high-quality adversarial attacks. Experimental results on benchmark dataset reveal that SOTA ADD systems exhibit significant vulnerabilities, with accuracies dropping from 98% to 26%, 92% to 54%, and 94% to 84% in white-box, gray-box, and black-box scenarios, respectively. When tested in other data sets, performance drops of 91% to 46%, and 94% to 67% were observed against the In-the-Wild and WaveFake data sets, respectively. These results highlight the significant vulnerabilities of existing ADD systems and emphasize the need to enhance their robustness against advanced adversarial threats to ensure security and reliability.*

## 1. Introduction

Audio deepfakes refer to synthetic audio clips created using advanced AI algorithms with the intent to deceive verification systems, *i.e.* automatic speaker verification (ASV), for potentially malicious purposes. Although audio deepfakes offer promising applications in various fields, (*e.g.* realistic voiceover), they also pose significant ethical and security challenges [4]. For example, malicious actors can exploit audio deepfakes for impersonation, cloning voices to spread misinformation, damage reputations, or commit financial fraud. These concerns highlight the need for solutions to counter the audio deepfakes.

In response, the research community has proposed var-ious audio deepfake detection (ADD) solutions to detect and mitigate the risks of audio deepfakes. ADD solutions can be categorized into two groups as shown in Figure 1. The first category includes non transcription based end-to-end ADD systems *i.e.* [8, 10, 23] and traditional ADD systems that use front-end features and backend classifiers *i.e.* LFCC [17], MFCC [6], and STDC [15]. The second category comprises transcription-based ADD methods that enhance performance against audio deepfakes by performing text matching [20]. These approaches have demonstrated impressive performance on benchmark datasets, such as ASVspoof2019 [25], showcasing their ability to detect a variety of deepfake audios.

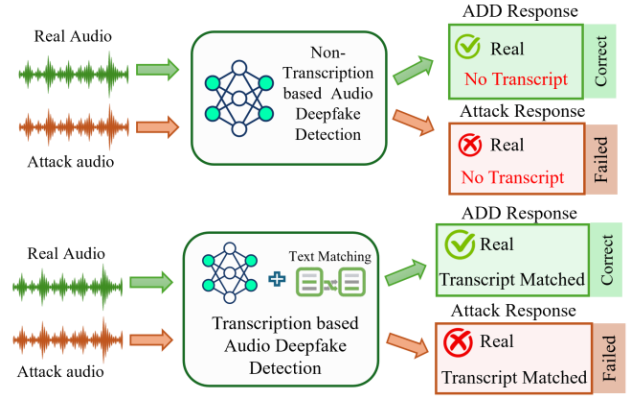

Figure 1. Comparison of audio deepfake detection (ADD) systems under adversarial attack: The upper block shows conventional ADD, while the lower block illustrates transcription-based ADD using transcript matching. The proposed attack bypasses both systems by aligning transcripts with input audios, highlighting their vulnerability.

Despite the impressive performance of ADD systems on benchmark dataset, state-of-the-art (SOTA) systems face critical challenges against adversarial attacks [16]. Most recent ADD systems that rely on deep learning solutions are more vulnerable to adversarial attacks. While recent studies show some trend to explore generalization across synthetic algorithms and datasets, they often overlook the risk of adversarial attacks. Adversarial attacks, typically crafted using advanced AI techniques *i.e.* generative adversarial net-works (GANs), can mislead ADD systems and compromise their reliability, as depicted in Figure 1. These attacks are classified based on the attacker's knowledge of the target model. In white-box attacks, the attacker has full access to the model's architecture, parameters, and training data, enabling highly effective attacks. Gray-box attacks occur when the attacker has partial knowledge, i.e., the architecture but not the parameters. In black-box attacks, the attacker has no internal knowledge of the model and relies solely on input-output queries, representing realistic real-world scenarios. Additionally, transferable adversarial attacks involve adversarial examples crafted for one model deceiving other models, highlighting shared vulnerabilities and posing significant risks across diverse ADD systems.

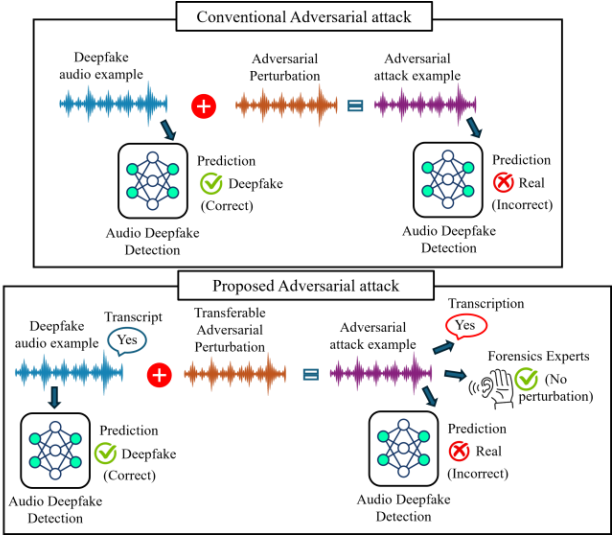

Figure 2. Comparison of traditional and proposed adversarial attacks in audio deepfake scenarios: (Top) Traditional adversarial attacks focus on inserting adversarial perturbations to deceive the model. (Bottom) Proposed adversarial attacks incorporate transferable perturbations while preserving transcription and perceptual integrity, deceiving both the model and forensic experts.

For example, Rabhi et al. [20] examined the robustness of ADD systems against traditional manipulations but does not include the challenges posed by GAN-based adversarial attacks. Afterward, in a recent study [2] the author tested ADD systems against two types of adversarial attacks based on GAN and proposed a transcription-based text matching solution to counter them. Given the advancements in both SOTA ADD methods and adversarial attacks, several key research questions arise that need to be addressed:
*(RQ1)*: Do SOTA ADD systems are generalize to the adversarial attacks?
*(RQ2)*: Are SOTA ADD systems effective against transferable adversarial attacks?
*(RQ3)*: How do SOTA ADD systems perform under white-box, gray-box, and black-box adversarial attack scenarios?

To address this critical gap and answer the aforementioned research questions, this paper proposes a novel, transferable adversarial attack to investigate the effectiveness of SOTA ADD methods. Unlike existing adversarial attacks [20,35], which rely on traditional methods to induce perturbations without considering transcription integrity or perceptual similarity as shown in Figure 2 (top), our approach ensures the preservation of both. This dual preservation enables the generation of high-quality, and transferable adversarial attacks that closely resemble real-world attack scenarios, as illustrated in Figure 2 (bottom).

Furthermore, we enhance the transferability and generalizability of the attack by leveraging an ensemble of surrogate models. Through this, we assess the vulnerability of SOTA ADD systems and demonstrate that the proposed adversarial attacks can effectively bypass ADD systems.

The main contributions of this paper are as follows:

- We propose a transferable adversarial attack that preserves transcription and perceptual integrity, making it vulnerable to existing ADD systems.

- We use diverse surrogate models to enhance the attack's transferability and generalizability, evaluating SOTA ADD systems across white-box, gray-box, and black-box scenarios.

- We validate the similarity of the attack to the input audio through qualitative and quantitative analyzes, ensuring real-world applicability.

- We conducted extensive experimental analysis to evaluate the performance of five state-of-the-art (SOTA) ADD systems against the proposed attack on three benchmark datasets: ASVspoof2019, WaveFake, and In-the-Wild.

The paper is organized as follows: Section 2 reviews related work, Section 3 details the proposed adversarial attack framework, Section 4 covers experiments and performance analysis, and Section 5 provides the conclusion of the study.

## 2. Related Works

This section provides an overview of existing ADD systems and adversarial attacks. Although research on adversarial attacks in ADD is limited, the relevant studies are discussed in the following subsections.

### 2.1. Audio Deepfake Detection

Audio deepfake detection methods can be broadly categorized into two main approaches: traditional handcrafted feature-based methods and modern end-to-end systems. The first category involves extracting critical features from audio signals using techniques such as

MFCCs [6], LFCCs [17], and Constant Q Cepstral Coefficients (CQCCs) [24]. This approach typically consists of two stages: feature extraction from raw audio as a frontend and classification using machine learning or deep learning algorithms as the backend. For example, studies such as [6, 17] have successfully employed MFCC and LFCC features, while others recent studies have explored advanced features *e.g*. Spectral Temporal Deviated Coefficients (STDC) [14] with transformer architecture and spectral contrast, envelop and mel spectrograms with siamese networks using one shot learning [13]. In recent, Khan et al. [16] evaluated 14 handcrafted features, such as RFCCs and SCMCs, across four classifiers (GMM, SVM, CNN, and CNN-GRU). The results showed that performance depends heavily on the extraction method, classifier, and dataset. For example, SCMCs performed best with GMM but poorly with others, while CQCCs showed dataset-specific variability, excelling on VSDC but underperforming on ASVSpoof2019.

In contrast, end-to-end ADD systems process raw audio or spectrograms for classification, leveraging advanced pattern recognition to detect fake audio. For instance, RawNet2 [23] uses sinc-layers to extract waveform features, ASSIST [10] employs graph attention networks for spectro-temporal relationships, and TSSDNet [8] utilizes a lightweight ResNet-based architecture for embeddings. Recent studies [3,19,22,31,40] highlight the potential of these methods. Unlike hand-crafted approaches, end-to-end systems offer a seamless, adaptable solution, prompting us to adopt them for evaluating ADD systems against our transferable adversarial attacks.

### 2.2. Adversarial Attacks

Adversarial attacks involve generating attack samples designed to mislead forensic methods. Various approaches have been developed to introduce adversarial attacks, ranging from traditional techniques to GAN-based adversarial methods. In the visual domain, these adversarial attacks [28] have been extensively explored to misguide deepfake image and video detection systems. However, the audio domain has seen comparatively limited research in this area, with only a few proposed methods focusing on either traditional [35] or advanced [20] adversarial attacks. Traditional adversarial attacks mainly alter the statistical properties of audio signals, such as fading, resampling, noise injection, and time-shifting [35]. For instance, Wu et al. [35] evaluated manipulation-based attacks, such as noise injection, volume control, fading, time stretching, resampling, time-shifting, and echo addition, against ADD models like RawNet2 [23],

In contrast, adversarial attacks aim to inject adversarial perturbations to deceive forensic methods. Adversarial attacks, particularly GAN-based adversarial attacks, can be accomplished in either a conventional [20] or trans-ferable [27] fashion. Zhang et al. [38] introduced the DeepFakeVox-HQ dataset to evaluate the robustness of existing ADD systems, including RawNet2 [23], RawGAT-ST [23], TE-ResNet [39], and RawNet3 [9], under realistic corruptions and adversarial attacks. The authors proposed robust training approaches for detecting fake audio, identifying adversarial attacks, and handling corrupted samples. Similarly, Kawa et al. [11] investigated adversarial defenses through superficial changes to input data, testing white-box and transferability attacks on models *i.e*. LCNN [36], SpecRNet [12], and RawNet3 [9]. The results demonstrated a decline in performance under adversarial attacks and introduced a novel adversarial training method that generalized well and resisted overfitting. However, this study ignored gray-box and black-box attack scenarios. Recently, Rabhi et al. [20] explored GAN-based adversarial attacks but limited their evaluation to a single victim model, Deep4SNet [2], and two attack scenarios.

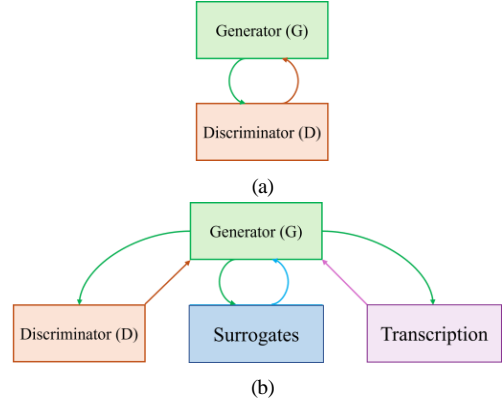

Figure 3. Illustration of adversarial attack approaches: (a) traditional GAN-based approach using a generator (G) and discriminator (D) to produce attack samples, (b) enhanced proposed approach integrating transcription modules with ADD and discriminator models to generate adversarial attacks.

This study evaluates the effectiveness of existing ADD systems against three adversarial threats, highlighting their vulnerabilities and limitations rather than proposing new defenses.

## 3. Proposed Work

This section presents the motivation, baseline analysis, architecture, and loss function implementation for the proposed transferable adversarial attack.

### 3.1. Motivation

Despite progress, the resilience of ADD systems against adversarial attacks remains an underexplored area. Existing research has primarily focused on conventional adversarial attacks, which often manipulate audio through noise addition, compression, and filtering. However, these traditional techniques tend to distort the audio content in unintended ways, diminishing their effectiveness. Furthermore, the impact of transferable adversarial attacks on ADD systems has been largely overlooked.

To overcome these limitations, we propose a novel transferable adversarial attack framework that preserves both transcription and perceptual integrity. By leveraging SOTA ADD systems and incorporating transcription information, our approach generates sophisticated adversarial examples that closely mimic real-world attack scenarios. This work not only exposes vulnerabilities in current ADD systems but also aims to advance the development of more robust detection methods capable of effectively identifying such adversarial threats.

### 3.2. GAN-based Adversarial Attack

As mentioned earlier, several adversarial attacks have been introduced in other domains, particularly in visual deepfake detection [26]. However, very few methods have explored adversarial attacks in the audio domain. In this section, we mainly focus on GAN-based adversarial attacks. Generally, GAN-based adversarial attacks fall into two categories: conventional [29] and transferable [41], as illustrated in Fig. 3. Conventional attacks follow a mini-max game theory, minimizing the distance between input and attacked samples, as shown in Fig. 3a. The generator model competes with the discriminator model to generate more realistic samples, defined as follows:

$$\min_G \max_D V(D, G) = E_{a \sim P_{real(a)}}[\log D(a)] + E_{a' \sim P_{deepfake(a')}}[1 - \log D(G(a'))] \quad (1)$$

where, $G$ and $D$ indicate generator and discriminator models, receptively. $a \sim P_{real(a)}$ and $a' \sim P_{deepfake(a')}$ represent the probability distribution of real and deepfake audios, receptively. $E_{a \sim P_{real(a)}}[\log D(a)]$ enables $D$ to correctly classify the real audio to maximize $D(a)$ while $E_{a' \sim P_{deepfake(a')}}[1 - \log D(G(a'))]$ allows $D$ to correctly classify deepfake audio to minimize $D(G(a'))$. $G(a')$ takes deepfake audio as input and injects adversarial attacks to generate attack audio and minimize the probability of $D$.

In contrast, transferable attacks are more effective and generalized to gray-box and black-box models by leveraging feedback from the discriminator and an ensemble of surrogate models. In the proposed method, we adopt a transferable attack to deceive the ADD methods by incorporating the discriminator, an ensemble of surrogate ADD models, and transcription models as shown in Fig. 3b, and defined as follows:

$$\min_G \max_D V(D, G) = E_{a \sim P_{real(a)}}[\log D(a)] + E_{a' \sim P_{deepfake(a')}}[1 - \log D(G(a'))] + E_{a' \sim P_{deepfake(a')}}[1 - \log C(G(a'))] + E_{a' \sim P_{deepfake(a')}}[1 - \log T(G(a'))] \quad (2)$$

where $E_{a' \sim P_{deepfake(a')}}[1 - \log T(G(a'))]$ provides feedback to generator using pretrained transcription model ($T$) to generate attack while preserving both transcription and perceptual integrity.

### 3.3. Architecture of the Proposed Attack

Fig. 4 depicts the overall architecture of the proposed adversarial attacks in audio domain. It encompasses generator, discriminator, ensemble of surrogates, and transcription models. The generator model competes with discriminator, ensemble of surrogates, and transcription model to accomplish the proposed adversarial attack. The details of each models are illustrated in the following subsections.

#### 3.3.1 Generator

The goal of the generator is to reconstruct a high-fidelity version of the audio signal that aligns with the natural characteristics of unaltered audio. The architecture of the generator is designed to process audio signals while maintaining temporal and spectral coherence, as shown in Table 1. It consists of multiple ConvBlocks followed by a feature map reduction module. Each ConvBlock is constructed with two 3×3 convolutional layers with stride 1 and Swish activation, followed by a 1×1 convolutional layer with Swish activation. To reduce the feature maps and ensure the output matches the dimensions of the input audio, the feature map reduction module uses a 3×3 convolutional layer with stride 1 and Tanh activation. Additionally, a residual connection, scaled by a learnable parameter $\alpha$, integrates the original input with the generated output, ensuring the preservation of essential signal characteristics. Post-processing steps, such as high-pass filtering, remove low-frequency artifacts, ensuring the generated audio remains clean and realistic.

#### 3.3.2 Discriminator

The goal of the discriminator is to distinguish between unaltered audio signals and those processed by the generator. The discriminator begins with a constrained convolutional layer designed to learn low-level forensic features critical for detecting anomalies in audio signals. The architecture starts with four convolutional blocks, each comprising a convolutional layer followed by synchronized batch normalization, hyperbolic tangent activation, and pooling op-erations. These layers gradually reduce the temporal resolution while capturing high-level forensic features. Finally, the discriminator includes three fully connected layers that process the extracted features into a high-dimensional representation, followed by sigmoid activation to predict the probability of the input being an unaltered or generated audio signal. The detailed architecture of the proposed discriminator is shown in Table 2.

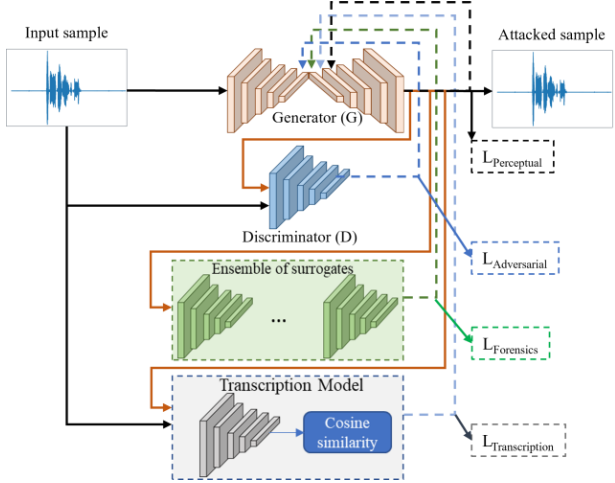

Figure 4. The architecture of the proposed transferable adversarial attack framework consisting of: an ensemble of surrogate models, a transcription model, and three associated loss functions. $A_{loss}$ the adversarial loss, $S_{loss}$ the surrogate loss, and $T_{loss}$ the transcription loss. The final loss is a weighted combination of these three losses. Solid lines indicate the forward pass, while dotted lines represent back-propagation for optimizing the model.

### 3.3.3 Ensemble of Surrogate Models

The goal of the surrogate forensic models for attack generation is to guide the generator in producing attacked audio samples that can fool forensic methods. Each surrogate model is pre-trained on the ASVspoof2019 dataset to perform audio deepfake detection. We select Res-

Table 1. Architecture of the proposed generator model.

| Type | Kernel | Shape | Activation |
|---|---|---|---|
| Input | - | (1, $L$) | - |
| Convolution ×2 | (3 × 3) | (64, $L$) | Swish |
| Convolution | (1 × 1) | (64, $L$) | Swish |
| Convolution ×2 | (3 × 3) | (128, $L$) | Swish |
| Convolution | (1 × 1) | (128, $L$) | Swish |
| Convolution ×2 | (3 × 3) | (256, $L$) | Swish |
| Convolution | (1 × 1) | (256, $L$) | Swish |
| Convolution ×2 | (3 × 3) | (128, $L$) | Swish |
| Convolution | (3 × 3) | (1, $L$) | Tanh |
| Residual Addition | - | (1, $L$) | - |
| High-pass Filtering | - | (1, $L$) | - |

Table 2. Architecture of the proposed discriminator model.

| Type | Kernel | Shape | Activation |
|---|---|---|---|
| Input | - | (1, $L$) | - |
| Constrained-CNN | (1 × 5) | (1, $L-4$) | - |
| Convolution ×2 | (7 × 7) | (64, $\frac{L-4}{2}$) | Tanh |
| Convolution ×2 | (5 × 5) | (64, $\frac{prev}{2}$) | Tanh |
| Convolution | (3 × 3) | (64, $\frac{prev}{2}$) | Tanh |
| Flatten | - | (47808) | - |
| Fully Connected | - | (256) | Tanh |
| Fully Connected | - | (128) | Tanh |
| Fully Connected | - | (1) | Sigmoid |

TSSDNet[1] [8], ResNet[2] [7] and Inc-TSSDNet [8] models as baseline surrogate models to train our generator for adversarial attack. To improve the transferability of the adversarial attack, we ensured diversity among the surrogate models. Specifically, three surrogates are chosen based on their distinct architectures and generalization capabilities to achieve a balanced trade-off between the attack success rate and the quality of the attacked audio. Furthermore, the selected models are widely adopted in this field, frequently used as baseline models in recent challenges [37, 40], and extensively studied in recent works [9,23,32]. Additionally, these models are publicly available with official implementations and pretrained weights. We conduct experiments using both two and three surrogate models as presented in Table 6. We evaluate the impact of increasing the number of surrogate models on attack performance and audio signal quality. Our results show that increasing from two to three surrogate models does not significantly improve attack performance or degrade audio signal quality. Considering this, we selected Res-TSSDNet and MS-ResNet as the final surrogate models for our experiments. These models are used as classifiers for white-box adversarial testing, utilizing the pre-trained weights provided by the authors to avoid training from scratch. For the gray-box scenario, we trained small and large variants of the surrogate models to assess their robustness against the proposed attack. In the black-box setting, we tested the effectiveness of our proposed attack using state-of-the-art ADD models, including RawNet2[3] [23], MS-ResNet[4] [33], and ResNet [7].

### 3.3.4 Transcription Models

The transcription model aims to preserve the integrity of the transcription in attacked audio. Subtle frequency changes introduced during the attack may alter the sounds of spoken words, leading to semantically different transcriptions. To mitigate this, we propose a transcription loss based on semantic similarity between the input and attacked audio. This loss ensures that the generator introduces forensic attacks without causing semantic-level transcription changes. Initially, we explored two self-supervised audio models, Wave2Vec [1] and Speech2Text [30]. Later, we selected the Wave2Vec model based on its superior transcription performance. Additionally, a transformer-based BERT encoder [21] was employed to generate text embeddings, with cosine similarity used to compute the transcription loss.

---
[1] https://github.com/ghua-ac/end-to-end-synthetic-speech-detection
[2] https://github.com/hsd1503/resnet1d
[3] https://github.com/Jungjee/RawNet?tab=readme-ov-file
[4] https://github.com/geekfeiw/Multi-Scale-1D-ResNet

### 3.4. Loss function

We train the generator and discriminator models while the ensemble of surrogates and transcription models remain frozen.

The generator model is trained with four losses to generate more realistic attack samples, defined as follows:

$$L_G = \lambda_1 L_{Perceptual} + \lambda_2 L_{Forensics} + \lambda_3 L_{Transcription} + \lambda_4 L_{Adversarial} \quad (3)$$

where $L_{Perceptual}$, $L_{Forensics}$, $L_{Transcription}$, and $L_{Adversarial}$ are the perceptual, forensics, transcription, and adversarial losses, respectively, and $\lambda_1$, $\lambda_2$, $\lambda_3$, and $\lambda_4$ are weights that balance the trade-off between quality and attack success rate, respectively.

The perceptual loss, $L_{Perceptual}$ is the $L_1$ loss between input and attack samples, defined as follows:

$$L_{Perceptual} = \frac{1}{S} \sum_{i=1}^{S} |A(i) - G(A'(i))| \quad (4)$$

where $S$ is the length of audio signal. $A$ and $A'$ indicate input and attack audio samples, respectively. The $L_{Perceptual}$ loss helps minimizing the distance between input and attack audio.

The forensics loss, $L_{Forensics}$ is cross-entropy classification loss between real and attack predictions, defined as follows:

$$L_{Forensics} = -\sum_{i \in F} \sum_{j \in N} \log F_i(G(A'_j)) \quad (5)$$

where $F$ and $N$ indicate the forensic methods and number of samples, respectively. $F_i$ is the $i-th$ forensic model of $F$. The $L_{Forensics}$ helps the generator to generate attacks in such a way so that it can alter the decision forensic methods.

The transcription loss, $L_{Transcription}$, is the contextual loss between the input and attacked audios. First, we transcribe the input and attacked audios to text using the *wav2vec* transcription model. Then, we compute two feature vectors, $f$ and $f'$, for the transcribed texts using a transformer model, defined as follows:

$$f = transformer(wav2vec(A)) \quad (6)$$

and

$$f' = transformer(wav2vec(A')) \quad (7)$$

Finally, we computed cosine-similarity between $f$ and $f'$, defined as follows:

$$L_{Transcription}(\mathbf{f}, \mathbf{f'}) = 1 - \frac{\mathbf{f} \cdot \mathbf{f'}}{\|\mathbf{f}\|\|\mathbf{f'}\|} \quad (8)$$

where,

$$\mathbf{f} \cdot \mathbf{f'} = \sum_{i=1}^{L} f_i f'_i, \quad \|\mathbf{f}\| = \sqrt{\sum_{i=1}^{L} f_i^2}, \quad \|\mathbf{f'}\| = \sqrt{\sum_{i=1}^{L} f'^2_i} \quad (9)$$

where $L$ indicates the dimension of the features $f_i$ and $f'_i$

The adversarial loss, $L_{Adversarial}$ is the binary cross-entropy loss for detecting attack audio by discriminator, defined as follows:

$$L_{Adversarial} = \log(1 - D_{\vartheta_D}(G(A'))) \quad (10)$$

The discriminator model is trained with binary cross entropy losses for discriminator real and fake predictions, defined as follows:

$$L_D = \log(1 - D(A)) + \log(1 - D(G(A'))) \quad (11)$$

## 4. Experimental Results

In this section, we explain the datasets, qualitative and quantitative attack analysis, and the performance of the selected baseline models in white-box, gray-box, and black-box scenarios.

### 4.1. Datasets

For the experimentation, we used the logical access (LA) subset of the benchmark ASVspoof2019 [34] dataset. The ASVspoof2019 dataset is commonly used as a standard dataset in the development of ADD systems. Furthermore, to assess the generalization of the attack and ADD system, we used the Wavefake [5] and In-the-wild [18] datasets.

### 4.2. Performance Evaluation and Comparisons

This section presents experiments to evaluate the performance of the proposed GAN-based adversarial attacks in white-box, gray-box, and black-box settings. We first assessed SOTA ADD methods, with results summarized in Table 3, showing their high effectiveness in distinguishing real and deepfake audio. Next, we analyzed their vulnerabilities under the adversarial attacks applied on benchmark audio deepfake datasets, with success rates summarized in Table 4. The following subsections discuss the observed performance degradation of the SOTA ADD methods in white-box, gray-box, and black-box settings.

Table 3. Baseline performance of the SOTA ADD methods.

| Eval. Type | Classifier | ASVspoof2019 | In-the-wild | WaveFake |
|---|---|---|---|---|
| White-box | Res-TSSDNet [8] | 98.46 | 95.46 | 97.46 |
| | Inc-TSSDNet [8] | 94.46 | 97.93 | 96.39 |
| Gray-box | Res-TSSDNet-S | 78.77 | 88.04 | 82.21 |
| | Res-TSSDNet-L | 92.09 | 96.32 | 96.60 |
| | Inc-TSSDNet-S | 92.22 | 67.33 | 91.78 |
| | Inc-TSSDNet-L | 93.27 | 95.55 | 99.76 |
| Black-box | RawNet2 [23] | 95.83 | 87.70 | 94.62 |
| | ResNet [7] | 99.14 | 98.15 | 99.68 |
| | MS-ResNet [33] | 94.67 | 94.75 | 99.06 |
| **Average** | - | 93.13 | 91.14 | 94.73 |

Table 4. Performance of the proposed attacks on the SOTA ADD methods.

| Eval. Type | Classifier | ASVspoof2019 | In-the-wild | WaveFake |
|---|---|---|---|---|
| White-box | Res-TSSDNet [8] | 26.33 | 0.38 | 0.42 |
| | Inc-TSSDNet [8] | 77.95 | 43.86 | 84.61 |
| Gray-box | Res-TSSDNet-S | 41.75 | 2.52 | 18.40 |
| | Res-TSSDNet-L | 84.20 | 64.08 | 93.47 |
| | Inc-TSSDNet-S | 54.40 | 15.02 | 79.71 |
| | Inc-TSSDNet-L | 84.18 | 75.20 | 95.13 |
| Black-box | RawNet2 [23] | 91.88 | 63.96 | 78.76 |
| | ResNet [7] | 85.91 | 62.88 | 63.54 |
| | MS-ResNet [33] | 84.48 | 90.46 | 89.20 |
| **Average** | - | 70.01 | 46.48 | 67.80 |

#### 4.2.1 White-box

We evaluate the effectiveness of the proposed attack in white-box settings, which represent a scenario where the attacker has full knowledge of the target ADD methods. Specifically, we assessed the attack's performance using the same surrogate models Res-TSSDNet and Inc-TSSDNet that were employed during the training phase for attack generation.

The performance degradation of Res-TSSDNet and Inc-TSSDNet is significantly more pronounced in white-box settings. As shown in Table 4, the performance drop for Res-TSSDNet is 72.13%, 95.08%, and 97.04%, while for Inc-TSSDNet, it is 16.51%, 54.07%, and 11.78% across the ASVspoof2019, In-the-Wild, and WaveFake datasets, respectively. The average performance drop is around 57% for white-box settings.

#### 4.2.2 Gray-box

We evaluate the performance of the proposed attack in gray-box settings, simulating an intermediate knowledge scenario where the attacker has partial understanding of the SOTA ADD methods being targeted. For this evaluation, we trained small and large variants of the surrogate models for the ADD task and utilized these models to conduct the gray-box analysis.

As shown in Table 4, the results indicate that the performance drop for the small variant of Res-TSSDNet under the adversarial attack is less severe compared to the original model. However, the large variant of Res-TSSDNet demonstrates greater robustness against the proposed adversarial attack. In the case of Inc-TSSDNet, the detection performance decreases for the small variant but improves for the large variant, suggesting that the proposed detection model may not have been adequately tested for adversarial robustness. In gray-box settings, the average performance degradation is about 30%.

#### 4.2.3 Black-box

We evaluate the performance of the proposed attack in black-box settings, simulating the worst-case scenario where the attacker has no knowledge of the SOTA ADD methods being targeted. For this black-box attack evaluation, we selected three SOTA ADD models: RawNet2, ResNet1D, and Multiscale-ResNet (MS-ResNet).

The evaluation results demonstrate the effectiveness of the proposed attack on all three models. However, RawNet2 shows greater robustness against the attack on the ASVspoof2019 dataset, with a performance drop of approximately 4%, but exhibits a more significant performance decline on the unseen datasets approximately 23.74% and 15.86% for In-the-wild and WaveFake, respectively. The performance drop for ResNet ranges from 14% to 28%, while MS-ResNet experiences a performance drop of only 4% to 10%, highlighting its greater robustness compared to ResNet.

### 4.3. Attack Analysis

This section presents both qualitative and quantitative analysis of the proposed attack, evaluating the quality of the attacks in terms of transcription accuracy and perceptual integrity.

#### 4.3.1 Qualitative Analysis

As mentioned earlier, the transcription-based approach enables the generation of high-quality attacks that minimize visual artifacts while preserving transcription accuracy and perceptual integrity. Figure 5 illustrates three attack samples, where the first, second, and third columns show the transcript texts, waveform, and spectrogram, respectively. The attack is evident in the transcript texts but remains visually undetectable in the waveform and spectrogram. The changes in the transcript before and after the attack are highlighted in yellow. The proposed attack focuses on making minimal alterations to the attack samples while successfully deceiving the SOTA ADD methods.

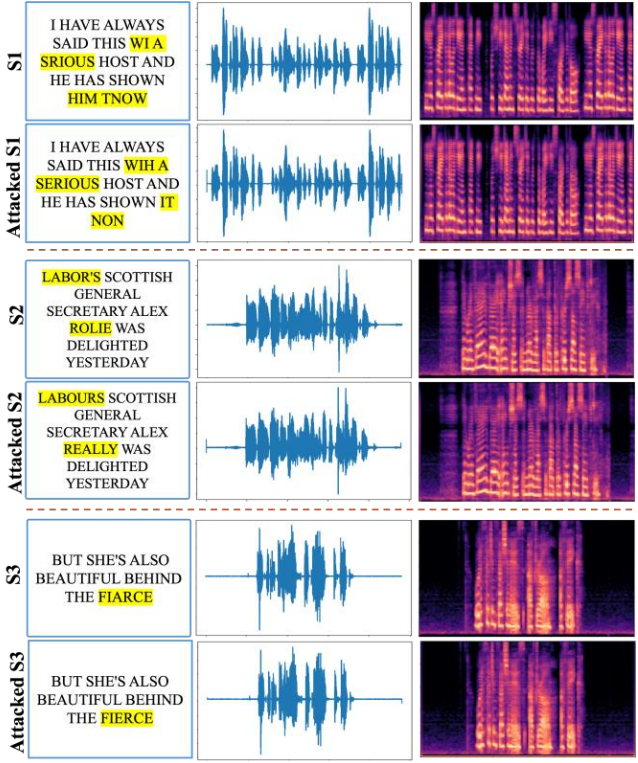

Figure 5. Visualization of fake samples labeled as S1, S2 and S3 and their corresponding attacked samples.

### 4.3.2 Quantitative Analysis

To provide concrete evidence, we report the quantitative results, including peak signal-to-noise ratio (PSNR), structural similarity index measure (SSIM), and text similarity for the attack samples. The proposed transferable adversarial attack is designed to effectively deceive the SOTA ADD methods without leaving visible attack signatures. To assess the visual quality of the audio signal, we compute the mean SSIM and PSNR between the original and attacked samples. To evaluate the preservation of transcription accuracy, we compare the text embedding similarity between the input and attacked samples using cosine similarity. The results, presented in Table 5, demonstrate that the proposed adversarial attack maintains both the quality of the audio and the accuracy of its transcription.

### 4.3.3 Effects of Surrogate and Balancing Parameters

To investigate the impact of the number of surrogate models and different balancing parameters, we conduct experiments using both two and three surrogate models, along with four different parameter settings. The results for PSNR, SSIM, baseline accuracy (BA), and attack accuracy (AA) are presented in Table 6. In contrast, reducing the surrogate balancing parameter from 0.1 to 0.0001 results in a substantial improvement in attack quality, with less performance degradation. Considering the trade-off between attack quality and performance drop, we select the 0.0001 parameter as it yields optimal results. Additionally, increasing the number of surrogate models from two to three does not significantly improve performance. However, increasing the number of surrogates raises the system's complexity. Therefore, the results presented in Table 4 use two surrogate models.

Table 5. Quantitative results of the proposed adversarial attack.

| Metric | ASVspoof2019 | In-the-wild | WaveFake |
|---|---|---|---|
| **PSNR** | 39.79 | 43.56 | 39.45 |
| **SSIM** | 0.99 | 0.98 | 0.96 |
| **Text Similarity** | 0.95 | 0.87 | 1.00 |

Table 6. Quantitative results of the proposed attacks on the ASVspoof2019 dataset with varying $\lambda$ values. Acc-BA and Acc-AA refer to the accuracies before and after the attack, respectively.

| $\lambda$ | PSNR | SSIM | Acc-BA | Acc-AA |
|---|---|---|---|---|
| | | 2 Surrogates | | |
| 0.1 | 16.72 | 0.54 | 97.34 | 49.29 |
| 0.01 | 17.37 | 0.74 | 97.34 | 50.18 |
| 0.001 | 30.08 | 0.89 | 97.34 | 60.42 |
| 0.0001 | **36.09** | **0.96** | 97.34 | **48.50** |
| | | 3 Surrogates | | |
| 0.1 | 16.49 | 0.57 | 97.34 | 45.64 |
| 0.01 | 16.75 | 0.69 | 97.34 | **38.39** |
| 0.001 | 16.67 | 0.98 | 97.34 | 61.88 |
| 0.0001 | **37.71** | **0.98** | 97.34 | 49.90 |

## 5. Conclusion

Transferable adversarial attacks are highly effective against deep neural network models but remain underexplored in audio deepfake detection (ADD). This paper addresses the gap by evaluating state-of-the-art (SOTA) ADD systems on three benchmark datasets using a novel transferable adversarial attack framework. Leveraging surrogate models, including self-supervised speech and text models, the framework preserves transcription and perceptual integrity. Experimental results show SOTA ADD systems fail against these attacks, with accuracy drops of 57%, 30.5%, and 6% in white-box, gray-box, and black-box scenarios, respectively, and up to 70%, 47%, and 68% on ASVspoof2019, In-the-Wild, and WaveFake datasets. These findings highlight the urgent need for robust ADD systems capable of resisting such threats.

## References

[1] Alexei Baevski, Yuhao Zhou, Abdelrahman Mohamed, and Michael Auli. wav2vec 2.0: A framework for self-supervised

learning of speech representations. *Advances in neural information processing systems*, 33:12449–12460, 2020. 6

[2] Dora M Ballesteros, Yohanna Rodriguez-Ortega, Diego Renza, and Gonzalo Arce. Deep4snet: deep learning for fake speech classification. *Expert Systems with Applications*, 184:115465, 2021. 2, 3

[3] Yujie Chen, Jiangyan Yi, Jun Xue, Chenglong Wang, Xiaohui Zhang, Shunbo Dong, Siding Zeng, Jianhua Tao, Lv Zhao, and Cunhang Fan. Rawbmamba: End-to-end bidirectional state space model for audio deepfake detection. *arXiv preprint arXiv:2406.06086*, 2024. 3

[4] Muhammad Umar Farooq, Awais Khan, Ijaz Ul Haq, and Khalid Mahmood Malik. Securing social media against deepfakes using identity, behavioral, and geometric signatures. *arXiv preprint arXiv:2412.05487*, 2024. 1

[5] Joel Frank and Lea Schönherr. Wavefake: A data set to facilitate audio deepfake detection. *arXiv preprint arXiv:2111.02813*, 2021. 6

[6] Ameer Hamza, Abdul Rehman Rehman Javed, Farkhund Iqbal, Natalia Kryvinska, Ahmad S Almadhor, Zunera Jalil, and Rouba Borghol. Deepfake audio detection via mfcc features using machine learning. *IEEE Access*, 10:134018–134028, 2022. 1, 3

[7] Shenda Hong, Yanbo Xu, Alind Khare, Satria Priambada, Kevin Maher, Alaa Aljiffry, Jimeng Sun, and Alexey Tumanov. Holmes: health online model ensemble serving for deep learning models in intensive care units. In *Proceedings of the 26th ACM SIGKDD International Conference on Knowledge Discovery & Data Mining*, pages 1614–1624, 2020. 5, 7

[8] Guang Hua, Andrew Beng Jin Teoh, and Haijian Zhang. Towards end-to-end synthetic speech detection. *IEEE Signal Processing Letters*, 28:1265–1269, 2021. 1, 3, 5, 7

[9] Jee-weon Jung, Hee-Soo Heo, Ju-ho Kim, Hye-jin Shim, and Ha-Jin Yu. Rawnet: Advanced end-to-end deep neural network using raw waveforms for text-independent speaker verification. *arXiv preprint arXiv:1904.08104*, 2019. 3, 5

[10] Jee-weon Jung, Hee-Soo Heo, Hemlata Tak, Hye-jin Shim, Joon Son Chung, Bong-Jin Lee, Ha-Jin Yu, and Nicholas Evans. Aasist: Audio anti-spoofing using integrated spectro-temporal graph attention networks. In *ICASSP 2022 - 2022 IEEE International Conference on Acoustics, Speech and Signal Processing (ICASSP)*, pages 6367–6371, 2022. 1, 3

[11] Piotr Kawa, Marcin Plata, and Piotr Syga. Defense against adversarial attacks on audio deepfake detection. *arXiv preprint arXiv:2212.14597*, 2022. 3

[12] Piotr Kawa, Marcin Plata, and Piotr Syga. Specrnet: Towards faster and more accessible audio deepfake detection. In *2022 IEEE International Conference on Trust, Security and Privacy in Computing and Communications (TrustCom)*, pages 792–799. IEEE, 2022. 3

[13] Awais Khan and Khalid Mahmood Malik. Securing voice biometrics: One-shot learning approach for audio deepfake detection. In *2023 IEEE International Workshop on Information Forensics and Security (WIFS)*, pages 1–6, 2023. 3

[14] Awais Khan and Khalid Mahmood Malik. Spotnet: A spoofing-aware transformer network for effective synthetic speech detection. In *Proceedings of the 2nd ACM International Workshop on Multimedia AI against Disinformation*, pages 10–18, 2023. 3

[15] Awais Khan, Khalid Mahmood Malik, and Shah Nawaz. Frame-to-utterance convergence: A spectra-temporal approach for unified spoofing detection. In *ICASSP 2024 - 2024 IEEE International Conference on Acoustics, Speech and Signal Processing (ICASSP)*, pages 10761–10765, 2024. 1

[16] Awais Khan, Khalid Mahmood Malik, James Ryan, and Mikul Saravanan. Battling voice spoofing: a review, comparative analysis, and generalizability evaluation of state-of-the-art voice spoofing counter measures. *Artificial Intelligence Review*, 56(Suppl 1):513–566, 2023. 1, 3

[17] Khaing Zar Mon, Kasorn Galajit, Candy Olivia Mawalim, Jessada Karnjana, Tsuyoshi Isshiki, and Pakinee Aimmanee. Spoof detection using voice contribution on lfcc features and resnet-34. In *2023 18th International Joint Symposium on Artificial Intelligence and Natural Language Processing (iSAI-NLP)*, pages 1–6. IEEE, 2023. 1, 3

[18] Nicolas M Müller, Pavel Czempin, Franziska Dieckmann, Adam Froghyar, and Konstantin Böttinger. Does audio deepfake detection generalize? *arXiv preprint arXiv:2203.16263*, 2022. 6

[19] Shah Nawaz, Muhammad Saad Saeed, Pietro Morerio, Arif Mahmood, Ignazio Gallo, Muhammad Haroon Yousaf, and Alessio Del Bue. Cross-modal speaker verification and recognition: A multilingual perspective. In *Proceedings of the IEEE/CVF conference on computer vision and pattern recognition*, pages 1682–1691, 2021. 3

[20] Mouna Rabhi, Spiridon Bakiras, and Roberto Di Pietro. Audio-deepfake detection: Adversarial attacks and countermeasures. *Expert Systems with Applications*, 250:123941, 2024. 1, 2, 3

[21] N Reimers. Sentence-bert: Sentence embeddings using siamese bert-networks. *arXiv preprint arXiv:1908.10084*, 2019. 6

[22] Saqlain Hussain Shah, Muhammad Saad Saeed, Shah Nawaz, and Muhammad Haroon Yousaf. Speaker recognition in realistic scenario using multimodal data. In *2023 3rd International Conference on Artificial Intelligence (ICAI)*, pages 209–213. IEEE, 2023. 3

[23] Hemlata Tak, Jose Patino, Massimiliano Todisco, Andreas Nautsch, Nicholas Evans, and Anthony Larcher. End-to-end anti-spoofing with rawnet2. In *ICASSP 2021-2021 IEEE International Conference on Acoustics, Speech and Signal Processing (ICASSP)*, pages 6369–6373. IEEE, 2021. 1, 3, 5, 7

[24] Massimiliano Todisco, Héctor Delgado, and Nicholas Evans. Constant q cepstral coefficients: A spoofing countermeasure for automatic speaker verification. *Computer Speech & Language*, 45:516–535, 2017. 3

[25] Massimiliano Todisco, Xin Wang, Ville Vestman, Md Sahidullah, Héctor Delgado, Andreas Nautsch, Junichi Yamagishi, Nicholas Evans, Tomi Kinnunen, and Kong Aik Lee. Asvspoof 2019: Future horizons in spoofed and fake audio detection. *arXiv preprint arXiv:1904.05441*, 2019. 1

[26] Kutub Uddin, Tae Hyun Jeong, and Byung Tae Oh. Counteract against gan-based attacks: A collaborative learning ap-

proach for anti-forensic detection. *Applied Soft Computing*, 153:111287, 2024. 4

[27] Kutub Uddin, Yoonmo Yang, Tae Hyun Jeong, and Byung Tae Oh. A robust open-set multi-instance learning for defending adversarial attacks in digital image. *IEEE Transactions on Information Forensics and Security*, 2023. 3

[28] Kutub Uddin, Yoonmo Yang, and Byung Tae Oh. Anti-forensic against double jpeg compression detection using adversarial generative network. In *Proceedings of the Korean Society of Broadcast Engineers Conference*, pages 58–60. The Korean Institute of Broadcast and Media Engineers, 2019. 3

[29] Kutub Uddin, Yoonmo Yang, and Byung Tae Oh. Analysis of generative adversarial network targeting anti-forensic in jpeg compressed domain. In *International Workshop on Advanced Imaging Technology (IWAIT) 2021*, volume 11766, pages 627–631. SPIE, 2021. 4

[30] Changhan Wang, Yun Tang, Xutai Ma, Anne Wu, Sravya Popuri, Dmytro Okhonko, and Juan Pino. Fairseq s2t: Fast speech-to-text modeling with fairseq. *arXiv preprint arXiv:2010.05171*, 2020. 6

[31] Chenglong Wang, Jiangyan Yi, Jianhua Tao, Haiyang Sun, Xun Chen, Zhengkun Tian, Haoxin Ma, Cunhang Fan, and Ruibo Fu. Fully automated end-to-end fake audio detection. In *Proceedings of the 1st International Workshop on Deepfake Detection for Audio Multimedia*, pages 27–33, 2022. 3

[32] Chenglong Wang, Jiangyan Yi, Jianhua Tao, Chuyuan Zhang, Shuai Zhang, Ruibo Fu, and Xun Chen. To-rawnet: improving rawnet with tcn and orthogonal regularization for fake audio detection. *arXiv preprint arXiv:2305.13701*, 2023. 5

[33] Fei Wang, Jinsong Han, Shiyuan Zhang, Xu He, and Dong Huang. Csi-net: Unified body characterization and action recognition. *arXiv preprint arXiv:1810.03064*, 2018. 5, 7

[34] Xin Wang, Junichi Yamagishi, Massimiliano Todisco, Héctor Delgado, Andreas Nautsch, Nicholas Evans, Md Sahidullah, Ville Vestman, Tomi Kinnunen, Kong Aik Lee, et al. Asvspoof 2019: A large-scale public database of synthesized, converted and replayed speech. *Computer Speech & Language*, 64:101114, 2020. 6

[35] Haolin Wu, Jing Chen, Ruiying Du, Cong Wu, Kun He, Xingcan Shang, Hao Ren, and Guowen Xu. Clad: Robust audio deepfake detection against manipulation attacks with contrastive learning. *arXiv preprint arXiv:2404.15854*, 2024. 2, 3

[36] Xiang Wu, Ran He, Zhenan Sun, and Tieniu Tan. A light cnn for deep face representation with noisy labels. *IEEE transactions on information forensics and security*, 13(11):2884–2896, 2018. 3

[37] Junichi Yamagishi, Xin Wang, Massimiliano Todisco, Md Sahidullah, Jose Patino, Andreas Nautsch, Xuechen Liu, Kong Aik Lee, Tomi Kinnunen, Nicholas Evans, et al. Asvspoof 2021: accelerating progress in spoofed and deepfake speech detection. In *ASVspoof 2021 Workshop-Automatic Speaker Verification and Spoofing Coutermeasures Challenge*, 2021. 5

[38] Zirui Zhang, Wei Hao, Aroon Sankoh, William Lin, Emanuel Mendiola-Ortiz, Junfeng Yang, and Chengzhi Mao. I can hear you: Selective robust training for deepfake audio detection. *arXiv preprint arXiv:2411.00121*, 2024. 3

[39] Zhenyu Zhang, Xiaowei Yi, and Xianfeng Zhao. Fake speech detection using residual network with transformer encoder. In *Proceedings of the 2021 ACM workshop on information hiding and multimedia security*, pages 13–22, 2021. 3

[40] Sheng Zhao, Qilong Yuan, Yibo Duan, and Zhuoyue Chen. An end-to-end multi-module audio deepfake generation system for add challenge 2023. *arXiv preprint arXiv:2307.00729*, 2023. 3, 5

[41] Xinwei Zhao, Chen Chen, and Matthew C Stamm. A transferable anti-forensic attack on forensic cnns using a generative adversarial network. *arXiv preprint arXiv:2101.09568*, 2021. 4